\documentclass[12pt,a4paper]{article}
\pdfoutput=1
\usepackage[T1]{fontenc}
\usepackage{color}
\usepackage{adjustbox}
\usepackage{amssymb,amsmath,bbold,MnSymbol}
\usepackage{epsf}
\usepackage{epsfig}
\usepackage{relsize}
\usepackage[dvipsnames]{xcolor}
\usepackage[linktoc=page,bookmarks=false,colorlinks=false,linkbordercolor=RoyalBlue,citebordercolor=ForestGreen,urlbordercolor=CornflowerBlue]{hyperref}
\usepackage{latexsym,mathrsfs,dsfont}
\usepackage[normalem]{ulem} % for strikeout \sout
\usepackage[compress]{cite}
\usepackage{graphicx}
\usepackage{url}
\usepackage{booktabs}
\usepackage{float}
\usepackage{multirow}
\usepackage{changepage}
\usepackage[hypcap]{caption, subcaption}

\usepackage[titles]{tocloft}

\usepackage{tabularx,colortbl}

\usepackage{dcolumn}  % format numbers in tabular to be aligned along . 

\usepackage{fancyhdr}
\advance \headheight by 3.0truept       % for 12pt mandatory...
\allowdisplaybreaks[1]

\definecolor{red}{cmyk}{0,1,1,0.4}
\definecolor{darkgreen}{rgb}{0.0,0.6,0.0}
\definecolor{cDarkGrey}{RGB}{91,91,91}
\definecolor{cGrey}{RGB}{245,243,238}
\definecolor{cBlue}{RGB}{0,110,191}
\definecolor{cLightBlue}{RGB}{214,237,252}
\definecolor{cRed}{RGB}{196,0,100}
\definecolor{cLightRed}{RGB}{254,222,237}
\definecolor{cGreen}{RGB}{0,166,80}
\definecolor{cLightGreen}{RGB}{254,222,237}
\definecolor{cOrange}{RGB}{221,74,44}
\definecolor{cLightOrange}{RGB}{255,215,210}
\definecolor{cPurple}{RGB}{93,35,125}
\definecolor{cLightPurple}{RGB}{241,230,252}
\definecolor{cYellow}{RGB}{252,191,10}
\definecolor{cISSRBlue}{RGB}{0,111,174}
\definecolor{cISSRGrey}{RGB}{167,169,172}

\newcommand{\be}{\begin{equation}}
\newcommand{\ee}{\end{equation}}

\newcounter{TODO}

\def \refsec#1{Section~\ref{#1}}

\def \refapp#1{Appendix~\ref{#1}}

\def \reftab#1{Table~\ref{#1}}

\DeclareMathOperator{\re}{Re}
\DeclareMathOperator{\im}{Im}

\newcommand{\bm}{\boldmath}

\newcommand{\oL}[1]{\overline{#1}}
\newcommand{\wT}[1]{\widetilde{#1}}

\newcommand{\MSbar}{${\overline{\text{MS}}}$}

\newcommand{\DF}{\Delta F}

\newcommand{\GeV}{\,\text{GeV}}
\newcommand{\geV}{\text{GeV}}
\newcommand{\TeV}{\,\text{TeV}}

\newcommand{\epe}{\varepsilon'/\varepsilon}

\newcommand{\epsK}{\varepsilon_K}

\newcommand{\alS}{\alpha_s}

\newcommand{\alE}{\alpha_\text{em}}

\newcommand{\muLow}{{\mu_\text{had}}}
\newcommand{\muEW}{{\mu_\text{ew}}}

\newcommand{\OpL}[2][{}]{Q_{#2}^{#1}}
\newcommand{\opL}[3][{}]{[Q_{#2}^{#1}]_{#3}}

\newcommand{\SUthreeC}{{\mathrm{SU(3)_c}}}

\newcommand{\Nf}{{N_f}}

\begin{document}

% \begin{flushleft}
% {\em Version of \today}
% \end{flushleft}

\vspace{-14mm}
\begin{flushright}
  {AJB-21-6}
\end{flushright}

\medskip

\begin{center}
{\Large\bf\bm
 BSM  Master Formula for $\epe$ in the WET Basis  \\[0.3cm]
  at NLO in QCD
}
\\[1.2cm]
{\bf
  Jason~Aebischer$^{a}$,
  Christoph~Bobeth$^{b}$,
  Andrzej~J.~Buras$^{b}$,
  Jacky Kumar$^{b}$
  }\\[0.5cm]

{\small
$^a$Department of Physics, University of California at San Diego,
    La Jolla, CA 92093, USA \\[0.2cm]
$^b$TUM Institute for Advanced Study,
    Lichtenbergstr. 2a, D-85747 Garching, Germany \\[0.2cm]
}
\end{center}

\vskip 1.0cm

\begin{abstract}
\noindent

As an important step towards a complete next-to-leading (NLO) QCD analysis
of the ratio $\epe$ within the Standard Model Effective Field Theory (SMEFT),
we present for the first time the NLO master formula for the BSM part of
this ratio expressed in terms of the Wilson coefficients of all contributing
operators evaluated at the electroweak scale. To this end we use the common
Weak Effective Theory (WET) basis (the so-called JMS basis) for which
tree-level and one-loop matching to the SMEFT are already known.
The relevant hadronic matrix elements of BSM operators at the electroweak
scale are taken from Dual QCD approach and the SM ones from lattice QCD.
It includes the renormalization group evolution and quark-flavour threshold
effects at NLO in QCD from hadronic scales, at which these matrix elements
have been calculated, to the electroweak scale.

\end{abstract}

\setcounter{page}{0}
\thispagestyle{empty}
\newpage

\setcounter{tocdepth}{2}
\setlength{\cftbeforesecskip}{0.21cm}

%\tableofcontents
%\newpage

%--------+---------+---------+---------+---------+---------+---------+---------+
%
%
%
%--------+---------+---------+---------+---------+---------+---------+---------+
\section{Introduction}

The ratio $\epe$ measures the size of the direct CP violation in $K_L\to\pi\pi$
decays $(\varepsilon^\prime)$ relative to the indirect one $(\varepsilon)$.
It is very sensitive to new sources of CP violation. As such it played a prominent
role in particle physics already for 45 years \cite{Buras:2021ane}. However,
due to large hadronic uncertainties in the evaluation of $\epe$ no consensus
between theorists has been reached on its value within the Standard Model (SM).
Even 20 years after its measurements  from NA48 \cite{Batley:2002gn} and KTeV
\cite{AlaviHarati:2002ye, Abouzaid:2010ny} collaborations implying the world
average
\begin{align}
  \label{EXP}
  (\epe)_\text{exp} & = (16.6\pm 2.3) \times 10^{-4} \,,
\end{align}
we do not know how much room is left for new physics (NP) contributions
to this ratio.

Within the SM $\epe$ is governed by the difference between QCD penguin (QCDP)
and electroweak penguin (EWP) contributions. Interestingly last year a consensus
between the lattice QCD (LQCD) RBC-UKQCD collaboration \cite{RBC:2020kdj}
and the Dual QCD (DQCD) estimate~\cite{Buras:2015xba} of the EWP contribution
to $\epe$ in the SM has been reached with the result
\be
  \label{EWPSM}
  (\epe)^{\text{EWP}}_\text{SM}
  = - (7 \pm 1) \times 10^{-4}
  \,,\qquad (\text{LQCD~and~DQCD}).
\ee
The estimate of this contribution from Chiral Perturbation Theory (ChPT),
having the same sign, is by roughly a factor of two lower in magnitude
\cite{Cirigliano:2019ani}, but in view of large uncertainties consistent
with \eqref{EWPSM}.

On the other hand there is no consensus on the size of the QCDP contribution,
which originates on the one hand in different estimates of the hadronic
matrix elements of QCDP operators and on the other hand in the estimate
of isospin-breaking effects that suppress the QCDP contributions. While the
RBC-UKQCD collaboration did not provide an estimate of the isospin-breaking
effects until now, the suppression of QCDP by them  provided by ChPT
\cite{Cirigliano:2019cpi} amounts to roughly $15\%$, while the advocates of
DQCD find the suppression as large as $30\%$ \cite{Buras:2020pjp}. This
difference can be traced to the $\eta-\eta^\prime$ mixing, which is included
in \cite{Buras:2020pjp} but not in \cite{Cirigliano:2019cpi}.

Adding isospin-breaking effects from \cite{Buras:2020pjp} to the RBC-UKQCD
results for QCDP hadronic matrix elements \cite{RBC:2020kdj} one finds
\cite{Buras:2020pjp, Aebischer:2020jto}
\begin{align}
  \label{SM}
  (\epe)_\text{SM} &
  = (13.9 \pm 5.2) \times 10^{-4} \,.
\end{align}
A similar result is found using ChPT \cite{Cirigliano:2019ani}, but as
discussed in \cite{Buras:2021ane} it is a numerical coincidence in view
of the smaller estimate of EWP contributions by these authors relative
to LQCD, and a different estimate of isospin-breaking corrections. In any
case, while this estimate is fully compatible with the experimental value
in \eqref{EXP}, the large uncertainty related primarily to QCDP contributions
does not exclude the modifications of $\epe$ by NP as large as $10^{-3}$.
In fact as argued on the basis of the DQCD approach \cite{Buras:2015xba}
in \cite{Buras:2021ane, Buras:2018ozh} the values of $\epe$ in the SM in
the ballpark of $5\times 10^{-4}$ are still possible.

This situation motivated various authors already for many years to perform
 analyses of $\epe$ in various extensions of the SM with the goal to
identify which models could allow for significant contributions to $\epe$
taking existing constraints from other processes into account. They are
listed in \cite{Aebischer:2019mtr}. However, one could question the usefulness
of these studies in view of very large uncertainties in the estimate of the
SM value of $\epe$ and thereby of the room left for NP contributions.
In particular one could question the usefulness of sophisticated NLO
calculations of Wilson coefficients of various operators within extensions
of the SM at present that are multiplied by very uncertain hadronic matrix
elements.

Here comes a good news. We have seen that the EWP contribution could be
determined within the SM with a respectable precision already now. Being
dominated by the $\Delta I=3/2$ transitions and contributing dominantly
to the isospin amplitude $A_2$, its contribution to $\epe$ turns out to
be automatically enhanced by a factor of 22.4 relative to the QCDP
contribution that dominates the $A_0$ amplitude. The only reason why in
the SM the QCDP contribution is, despite of this strong suppression
relative to $A_2$, larger in magnitude than the EWP contribution is that
at scales relevant to the $K_L\to\pi\pi$ decays the value of $\alS$ is by
roughly a factor of 50 larger than the one of $\alE$. Consequently the Wilson
coefficients of QCDP operators are much larger than those of EWP
operators and this implies a positive sign of $\epe$ as opposed to the
result in \eqref{EWPSM}.

While one cannot exclude that such a pattern of different contributions
is present in some of the NP scenarios, in most of the BSM analyses to date
the contributions of NP to the imaginary part of $A_2$ were by far dominant.
It could be then that BSM contributions to $\epe$, as far as hadronic
uncertainties are concerned, can be more precisely calculated than the SM
contribution. Therefore, as proposed in \cite{Buras:2015jaq}, it is useful to
write $\epe$ as a sum of the SM and BSM contributions,
\begin{align}
  \label{AJBeprime}
  \frac{\varepsilon'}{\varepsilon} &
  = \left(\frac{\varepsilon'}{\varepsilon}\right)_\text{SM}
  + \left(\frac{\varepsilon'}{\varepsilon}\right)_\text{BSM}
  \,,
\end{align}
and concentrate on the BSM one, until an improved estimate of the SM
contribution will be available.

Now, the novelty of the BSM contributions to $\epe$ is the presence of new
operators, absent or strongly suppressed in the SM. The hadronic matrix
elements of all BSM four-quark operators have been calculated to date only
within DQCD in \cite{Aebischer:2018rrz} and it will still take some time
before we know LQCD estimates for them. But already this estimate allowed
to derive a master formula for $\epe$ in \cite{Aebischer:2018quc}.
Knowing the Wilson coefficients in any BSM scenario at the electroweak scale,
this formula  provides the BSM contribution to $\epe$.
This then allowed to perform for the first time a SMEFT anatomy of $\epe$
\cite{Aebischer:2018csl}, demonstrating the possible interplay of NP contributions to $\epe$ with BSM effects in other observables.

So far the master formula for $\epe$ within WET in \cite{Aebischer:2018quc}
is based on LO RG evolution from the low-energy scale $\muLow$ to
the electroweak scale $\muEW$, at which the Wilson coefficients (WCs)
are obtained from a UV completion or SMEFT. However, beyond the LO,
the RG evolution from $\muLow$ to $\muEW$ is known only in the so-called
\emph{BMU basis} \cite{Buras:2000if}, which is less suitable for the matching 
to the SMEFT. Moreover, for this matching the most convenient WET basis is
the one introduced
in \cite{Jenkins:2017jig}, the so-called \emph{JMS basis}.\footnote{We use
here the \texttt{WCxf} convention for the JMS basis, defined in
\cite{Aebischer:2017ugx}.} In particular, for this basis the tree-level
matching of SMEFT on to WET \cite{Jenkins:2017jig} and the corresponding
one-loop matching \cite{Dekens:2019ept, Aebischer:2015fzz} are known.
Therefore it is evident that a master formula for $\epe$ written in terms
of WCs in the JMS basis would be more useful for BSM analyses than the one
that uses BMU basis at the electroweak scale. Moreover, for future combination
of WCs with hadronic matrix elements from LQCD it is mandatory to calculate
the RG evolution at the NLO level.

With this goal in mind we have recently calculated, including NLO QCD
corrections, the RG evolution matrix $\hat U_\text{BMU}(\muLow,\, \muEW)$
\cite{Aebischer:2021raf} that allows to express the WCs
$\vec{C}_\text{BMU}(\muLow)$ at the hadronic scale in the BMU basis
in terms of the $\vec{C}_\text{JMS}(\muEW)$ in the JMS basis at the
electroweak scale as follows
\begin{align}
  \label{ABBKM}
  \vec{C}_\text{BMU}(\muLow)
  \; = \; \hat U_\text{BMU}(\muLow, \muEW) \;
  \hat M_\text{JMS}(\muEW)\; \vec{C}_\text{JMS}(\muEW) .
\end{align}
To this end the RG evolution from $\muLow$ to $\muEW$ has been performed
in the BMU basis, including quark-flavour threshold crossings at NLO
in QCD, and subsequently the resulting $\vec{C}_\text{BMU}(\muEW)$
in the BMU basis have been transformed to $\vec{C}_\text{JMS}(\muEW)$
in the JMS basis at the one-loop level. This required a careful treatment
of the evanescent operators, encoded in the matrix $\hat M_\text{JMS}(\muEW)$.

Having the matrices $\hat U_\text{BMU}$ and $\hat M_\text{JMS}$ at our disposal,
the main result of our work is a new master formula for $(\epe)_\text{BSM}$,
which is superior to the one in \cite{Aebischer:2018quc, Aebischer:2018csl,
Aebischer:2020jto} in two ways
\begin{itemize}
\item
  It includes NLO QCD RG evolution below the electroweak scale.
  Further, also the LO QED and NLO QED$\times$QCD evolution is known,
  and can be applied if needed \cite{Aebischer:2021raf}.
\item
  It is expressed in terms of the WCs evaluated at the electroweak scale
  in the JMS basis, thereby allowing with the help of the results in
  \cite{Dekens:2019ept} to express it eventually in terms of the SMEFT WCs
  including one-loop corrections in the matching of the SMEFT on to WET.
  As such it can be readily used for all UV completions that reduce
  to SMEFT in their infra-red regime.
\end{itemize}

Optimally for model building it would be useful to have a master formula
for $\epe$ expressed in terms of WCs in the Warsaw SMEFT basis that are
evaluated at the NP scale as we have done for $\Delta F=2$ transitions in
\cite{Aebischer:2020dsw}. We will present such a formula in a future
publication.

As the two items above have been accomplished in \cite{Aebischer:2021raf},
the main goal of the present paper is the evaluation of all
hadronic matrix elements for $I=0$ and $I=2$ in the BMU basis at $\muLow$
using the results of \cite{Aebischer:2018rrz} obtained in the DQCD approach.
Combining them with \eqref{ABBKM}, we will be able to present a new master
formula for $\epe$ with the virtues listed above.

Our paper is organized as follows. In \refsec{sec:2} we present the general
expression for the updated master formula for $\epe$. In \refsec{sec:3} we
present the numerical values of the relevant hadronic matrix elements and
in \refsec{sec:4} the values of the coefficients $P_b(\muEW)$ for the
40 WCs in the JMS basis will be presented. The summary of our results and
an outlook are presented in \refsec{sec:5}. The definitions of the
JMS and BMU operator bases are given in \refapp{app:def-JMS} and
\refapp{app:def-BMU}, respectively.

%--------+---------+---------+---------+---------+---------+---------+---------+
%
%
%
%--------+---------+---------+---------+---------+---------+---------+---------+
\section[Master formula for $\epe$]
{\bm Master formula for $\epe$}
\label{sec:2}

The general formula for $\epe$ reads \cite{Buras:2015yba}
\begin{align}
  \label{eq:epe-formula}
  \frac{\varepsilon'}{\varepsilon} &
  = -\,\frac{\omega_+}{\sqrt{2}\,|\varepsilon_K|} \left[\,
      \frac{\im A_0}{\re A_0}\,(1 - \hat\Omega_\text{eff})
    - \frac{1}{a}\frac{\im A_2}{\re A_2} \,\right],
\end{align}
with $\omega_+$, $a$ and $\hat\Omega_\text{eff}$ given
as follows
\begin{align}
  \label{OM+}
  \omega_+ & = a\,\frac{\re A_2}{\re A_0} = (4.53 \pm 0.02) \times 10^{-2},
&
  a & = 1.017,
&
  \hat\Omega_\text{eff} & = (29 \pm 7) \times 10^{-2} \,.
\end{align}
Here $a$  and $\hat\Omega_\text{eff}$ summarize isospin-breaking corrections
and include strong isospin violation $(m_u \neq m_d)$, the correction to the
isospin limit coming from $\Delta I=5/2$ transitions and electromagnetic
corrections \cite{Cirigliano:2003nn, Cirigliano:2003gt, Bijnens:2004ai}.
The value for $\hat\Omega_\text{eff}$ quoted above is the most recent one
that includes the effect of $\eta-\eta^\prime$ mixing \cite{Buras:2020pjp}.
Without it one finds the value $\hat\Omega_\text{eff} = (17\pm 9) \times
10^{-2}$ \cite{Cirigliano:2019cpi}.

In the SM $\im A_0$ receives dominantly contributions from QCDPs, but also
from EWP. On the other hand, $\im A_2$ receives contributions exclusively
from EWPs. However, in BSM scenarios they are both affected by contributions
of BSM operators and also the WCs of SM operators are modified.
The amplitudes $\re A_{0,2}$ are extracted from the branching ratios of
$K\to\pi\pi$ decays in the isospin limit. Their values are given by
\begin{align}
  \label{eq:6.3}
  \re A_0 & = 27.04(1)\times 10^{-8}\GeV \, ,
&
  \re A_2 & = 1.210(2)\times 10^{-8}\GeV \, .
\end{align}
The main reason for taking these amplitudes from the data is that we do
not know presently the BSM contributions to them and proceeding in this
manner one automatically includes both SM and BSM contributions.

For the discussion of the BSM contributions with the WCs normalized at the
electroweak scale we define
\begin{align}
  \label{eq:abbs}
  {\mathcal H}_{\Delta S=1} & \equiv -\sum_b \Big[
    C_b(\muEW)\, \OpL{b} + C_b^\prime(\muEW)\, \OpL[\prime]{b} \Big],
\end{align}
with the operators $\OpL{b}$ and $\OpL[\prime]{b}$ given in the JMS basis
as defined in \refapp{app:def-JMS} and the ordering of operators in
VLL, VLR and SRR sectors as given in \cite{Aebischer:2021raf}.
The chirality-flipped operators $\OpL[\prime]{b}$ in the sectors VRR,
VRL and SLL are obtained from $\OpL{b}$ by interchanging the
chirality-projectors $P_L\leftrightarrow P_R$.
The relation of $C^\prime_a(\muLow)$ in the BMU basis to $C^\prime_b(\muEW)$
in the JMS basis is given in \eqref{ABBKM}.

The updated master formula \cite{Aebischer:2018quc} for the BSM part in
\eqref{AJBeprime} reads
\begin{align}
  \label{eq:master}
  \left(\frac{\varepsilon'}{\varepsilon}\right)_\text{BSM} &
  = \;\; \sum_b  P_b(\muEW)
    \im \big[ C_b(\muEW) - C^\prime_b(\muEW) \big]
    \times (1 \TeV)^2,
\end{align}
where
\begin{align}
  \label{eq:master2}
  P_b(\muEW) &
  = \sum_{a} \sum_{I=0,2} \;p_{ba}^{(I)}(\muEW,\, \muLow)
  \,\left[\frac{\langle \OpL{a}(\muLow) \rangle_I}{\geV^3}\right],
\end{align}
with the sum over $b$ running over the Wilson coefficients $C_b$ of all
operators in the JMS basis and their chirality-flipped counterparts
denoted by $C_b'$. The relative minus sign accounts for the fact that
their $K\to\pi\pi$ matrix elements differ by a sign. Among the contributing
operators are also operators present already in the SM, but their WCs in \eqref{eq:master} are meant to include only BSM contributions.
The list of all contributing operators in the JMS basis is given in
\refapp{app:def-JMS} and the one for the BMU basis in \refapp{app:def-BMU}.
We mention further that the choice of the JMS basis in
\cite{Aebischer:2021raf} contains hermitian conjugated operators
$\opL[V1,LR]{uddu}{1211}^\dagger,\; \opL[V8,LR]{uddu}{1211}^\dagger,
\; \opL[V1,LR]{uddu}{2212}^\dagger$ and $\opL[V8,LR]{uddu}{2212}^\dagger$
in \eqref{eq:abbs}, such that actually their complex-conjugated Wilson
coefficients enter $\epe$, but using $\im C_b(\muEW)^* = -\im C_b(\muEW)$,
in~\eqref{eq:master} the minus sign is absorbed into $P_b$.

It should be emphasized that the BMU operators
\begin{align}
  \OpL{i},  \qquad  i=19-22, \quad  i=25-28, \quad i=37-40 ,
\end{align}
can not be generated in the tree-level matching from SMEFT operators, because
they do not conserve hypercharge.

Comparing \eqref{eq:master} with \eqref{eq:epe-formula} and taking the
minus sign in \eqref{eq:abbs} into account we have\footnote{Note that
    on the l.h.s the indices are $ba$, while on the r.h.s $ab$.}
\begin{align}
   \label{eq:def-pba0}
   p_{ba}^{(0)}(\muEW, \muLow) &
   = \frac{\omega_+}{\sqrt{2} \, |\epsK| }
     \left[\frac{10^{-6} \GeV}{\re A_0} \right] \,
   (1-\hat\Omega_\text{eff}) \,
   [\hat U_\text{BMU}(\muLow,\muEW) \hat M_\text{JMS}(\muEW)]_{ab}\,,
\\
   \label{eq:def-pba2}
   p_{ba}^{(2)}(\muEW, \muLow) &
   = -\frac{\omega_+}{\sqrt{2} a \,|\epsK|} \,
     \left[\frac{10^{-6} \GeV}{\re A_2} \right]
   [\hat U_\text{BMU}(\muLow,\muEW) \hat M_\text{JMS}(\muEW)]_{ab}\,,
\intertext{leading to}
   p_{ba}^{(0)}(\muEW, \muLow) &
   = - a (1 - \hat\Omega_\text{eff})\, \frac{\re A_2}{\re A_0} \,
     p_{ba}^{(2)}(\muEW, \muLow)
\label{eq:pba0-pba2}
\end{align}
and therefore implying a strong suppression of $I=0$ coefficients relative
to the $I=2$ ones by ${\re A_2}/{\re A_0} \approx 1/22$. On the other hand,
this suppression is partially lifted because in most cases 
$\langle \OpL{a} \rangle_2 < \langle \OpL{a} \rangle_0$.

\begin{table}
\begin{center}
\begin{adjustbox}{width=1\textwidth}\begin{tabular}{lrrrrrrrr}
\toprule
      MEs & $[C_{ud}^{V1,LL}]_{1121}$ & $[C_{ud}^{V8,LL}]_{1121}$ & $[C_{ud}^{V1,LL}]_{2221}$ & $[C_{ud}^{V8,LL}]_{2221}$ & $[C_{dd}^{VLL}]_{2133}$ & $[C_{dd}^{VLL}]_{2331}$ & $[C_{dd}^{VLL}]_{2111}$ & $[C_{dd}^{VLL}]_{2122}$ \\
\midrule
    $Q_1$ &                   -1494.8 &                     554.0 &                    1494.8 &                    -554.0 &                       - &                       - &                       - &                       - \\
    $Q_2$ &                     609.7 &                    -849.0 &                    -609.7 &                     849.0 &                       - &                       - &                       - &                       - \\
    $Q_3$ &                      21.6 &                     -23.0 &                    -476.8 &                     161.6 &                  -282.5 &                  1092.4 &                  -368.8 &                  -368.8 \\
    $Q_4$ &                     -24.4 &                      44.9 &                     179.0 &                    -238.1 &                   286.9 &                 -1090.6 &                   114.6 &                   114.6 \\
    $Q_9$ &                         - &                         - &                    -996.5 &                     369.3 &                   291.8 &                 -1111.2 &                   352.4 &                   352.4 \\
 $Q_{10}$ &                         - &                         - &                     406.5 &                    -566.0 &                  -291.8 &                  1111.2 &                   -57.3 &                   -57.3 \\
 $Q_{14}$ &                         - &                         - &                         - &                         - &                       - &                       - &                  -442.5 &                   442.5 \\
    $Q_5$ &                      -6.7 &                      -4.1 &                      -6.3 &                      -5.1 &                    -5.5 &                     3.8 &                   -18.0 &                   -18.0 \\
    $Q_6$ &                     -32.2 &                      68.7 &                     -31.1 &                      66.9 &                   -13.5 &                    65.1 &                    90.2 &                    90.2 \\
\bottomrule
\end{tabular}
\end{adjustbox}                                            \begin{adjustbox}{width=1\textwidth}\begin{tabular}{lrrrrrrrr}
\toprule
      MEs & $[C_{du}^{V1,LR}]_{2111}$ & $[C_{du}^{V8,LR}]_{2111}$ & $[C_{du}^{V1,LR}]_{2122}$ & $[C_{du}^{V8,LR}]_{2122}$ & $[C_{dd}^{V1,LR}]_{2133}$ & $[C_{dd}^{V8,LR}]_{2133}$ & $[C_{dd}^{V1,LR}]_{2111}$ & $[C_{dd}^{V8,LR}]_{2111}$ \\
\midrule
    $Q_3$ &                     -19.4 &                     -27.1 &                     -26.2 &                     -39.2 &                     -10.3 &                     -19.2 &                     -19.4 &                     -27.1 \\
    $Q_4$ &                      18.1 &                      49.9 &                      38.4 &                      86.1 &                      10.3 &                      32.7 &                      18.1 &                      49.9 \\
    $Q_5$ &                    -345.6 &                       3.8 &                       1.1 &                     -17.3 &                       3.4 &                      -2.9 &                    -345.6 &                       3.8 \\
    $Q_6$ &                    -363.9 &                    -605.1 &                      52.7 &                     135.0 &                      19.2 &                      71.0 &                    -363.9 &                    -605.1 \\
    $Q_7$ &                    -707.0 &                      18.0 &                         - &                         - &                         - &                         - &                     353.5 &                      -9.0 \\
    $Q_8$ &                    -792.5 &                   -1407.8 &                         - &                         - &                         - &                         - &                     396.2 &                     703.9 \\
 $Q_{15}$ &                         - &                         - &                         - &                         - &                         - &                         - &                    -594.4 &                   -1055.8 \\
 $Q_{16}$ &                         - &                         - &                         - &                         - &                         - &                         - &                    -530.2 &                      13.5 \\
\bottomrule
\end{tabular}
\end{adjustbox}                                            \begin{adjustbox}{width=1\textwidth}\begin{tabular}{lrrrrrrrr}
\toprule
          MEs & $[C_{dd}^{V1,LR}]_{2122}$ & $[C_{dd}^{V8,LR}]_{2122}$ & $[C_{uddu}^{V1,LR}]_{1211}$ & $[C_{uddu}^{V8,LR}]_{1211}$ & $[C_{uddu}^{V1,LR}]_{2212}$ & $[C_{uddu}^{V8,LR}]_{2212}$ & $[C_{dd}^{V1,LR}]_{2331}$ & $[C_{dd}^{V8,LR}]_{2331}$ \\
\midrule
        $Q_3$ &                     -19.4 &                     -27.1 &                           - &                           - &                           - &                           - &                         - &                         - \\
        $Q_4$ &                      18.1 &                      49.9 &                           - &                           - &                           - &                           - &                         - &                         - \\
        $Q_5$ &                    -345.6 &                       3.8 &                           - &                           - &                           - &                           - &                         - &                         - \\
        $Q_6$ &                    -363.9 &                    -605.1 &                           - &                           - &                           - &                           - &                         - &                         - \\
        $Q_7$ &                     353.5 &                      -9.0 &                           - &                           - &                           - &                           - &                         - &                         - \\
        $Q_8$ &                     396.2 &                     703.9 &                           - &                           - &                           - &                           - &                         - &                         - \\
     $Q_{15}$ &                     594.4 &                    1055.8 &                           - &                           - &                           - &                           - &                         - &                         - \\
     $Q_{16}$ &                     530.2 &                     -13.5 &                           - &                           - &                           - &                           - &                         - &                         - \\
 $\rm Q_{19}$ &                         - &                         - &                      2120.9 &                       -53.9 &                           - &                           - &                         - &                         - \\
     $Q_{20}$ &                         - &                         - &                      2377.4 &                      4223.4 &                           - &                           - &                         - &                         - \\
\bottomrule
\end{tabular}
\end{adjustbox}                                            \begin{adjustbox}{width=1\textwidth}\begin{tabular}{lrrrrrrrr}
\toprule
      MEs & $[C_{dd}^{S1,RR}]_{2111}$ & $[C_{dd}^{S8,RR}]_{2111}$ & $[C_{dd}^{S1,RR}]_{2122}$ & $[C_{dd}^{S8,RR}]_{2122}$ & $[C_{ud}^{S1,RR}]_{1121}$ & $[C_{ud}^{S8,RR}]_{1121}$ & $[C_{uddu}^{S1,RR}]_{1121}$ & $[C_{uddu}^{S8,RR}]_{1121}$ \\
\midrule
 $Q_{25}$ &                   -2723.9 &                     922.2 &                         - &                         - &                         - &                         - &                           - &                           - \\
 $Q_{26}$ &                     -12.4 &                      30.3 &                         - &                         - &                         - &                         - &                           - &                           - \\
 $Q_{27}$ &                         - &                         - &                   -2723.9 &                     922.2 &                         - &                         - &                           - &                           - \\
 $Q_{28}$ &                         - &                         - &                     -12.4 &                      30.3 &                         - &                         - &                           - &                           - \\
 $Q_{29}$ &                         - &                         - &                         - &                         - &                    -695.0 &                    -693.0 &                      2176.2 &                        55.0 \\
 $Q_{30}$ &                         - &                         - &                         - &                         - &                   -5763.7 &                     -80.6 &                      3393.1 &                       927.7 \\
 $Q_{31}$ &                         - &                         - &                         - &                         - &                    -403.0 &                     -89.8 &                       733.3 &                        -2.8 \\
 $Q_{32}$ &                         - &                         - &                         - &                         - &                     111.0 &                      -0.9 &                      -138.3 &                        37.9 \\
\bottomrule
\end{tabular}
\end{adjustbox}   
%                                         \begin{adjustbox}{width=1\textwidth}\begin{tabular}{lrrrrrrrr}
%\toprule
%Empty DataFrame
%Columns: Index(['MEs', '$[C_{ud}^{S1,RR}]_{2221}$', '$[C_{ud}^{S8,RR}]_{2221}$',
%       '$[C_{uddu}^{S1,RR}]_{2122}$', '$[C_{uddu}^{S8,RR}]_{2122}$',
%       '$[C_{dd}^{S1,RR}]_{2133}$', '$[C_{dd}^{S8,RR}]_{2133}$',
%       '$[C_{dd}^{S1,RR}]_{2331}$', '$[C_{dd}^{S8,RR}]_{2331}$'],
%      dtype='object')
%Index: Int64Index([], dtype='int64') \\
%\bottomrule
%\end{tabular}
%\end{adjustbox}                                            

\end{center}
\caption{The values for $p_{ba}^{(2)}(\muEW, \mu_\text{had})$
  (see \eqref{eq:def-pba2} for definition) are shown. Here the first index $a$
  refers to the matrix elements in the BMU basis as shown in the left most
  column and $b$ is for the Wilson coefficients in the JMS basis shown in the
  top row of the table. The corresponding $p_{ba}^{(0)}$ can be obtained through
  the relation \eqref{eq:pba0-pba2}.}
\end{table}

So far only four-quark operators have been considered, however also the
chromo- and electromagnetic dipole operators ($\OpL{8g}$ and $\OpL{7\gamma}$
respectively) contribute to $\epe$ in principle. Whereas the electromagnetic
dipole operator has been neglected, the $\OpL{8g}$ was included in the previous
LO master formula \cite{Aebischer:2018quc, Aebischer:2018csl, Aebischer:2020jto}.
Also in this case an improvement towards NLO QCD evolution would be desirable,
but here the corresponding anomalous dimension matrices for the mixing of the
four-quark operators into $\OpL{8g}$ are not yet available, except the NLO
mixing of $\OpL{8g}$ and $\OpL{7\gamma}$ \cite{Misiak:1994zw}. In view
of this, we include contributions that are proportional to $\langle \OpL{8g}
\rangle_I$ at LO only, based on the results of \cite{Aebischer:2018csl}.
The involved operators belong to the three sectors SRR,$s,c,b$ (called
class B in \cite{Aebischer:2018quc}) and the sectors SRR$,u$ (class C) and SRR$,d$
(class D), which are part of the BMU basis. Eventually it remains to transform
their Wilson coefficients at $\muEW$ to the JMS basis, where we use the LO
transformations given in \cite{Aebischer:2021raf}, see also \refapp{app:SD2-BMU}.
In the JMS and BMU bases the chromomagnetic dipole operator are defined as
\begin{align}
  \opL[]{dG}{21} &
  = [\bar{s} \sigma^{\mu\nu} P_{R} T^A d] G^{a}_{\mu\nu} \,,
  & \text{(JMS)}
\\
  \OpL[\prime]{8g} &
  = m_s [\bar{s} \sigma^{\mu\nu} P_{R} T^A d] G^{a}_{\mu\nu} \,,
  & \text{(BMU)} &
\end{align}
respectively, which differ by a factor of the strange quark mass (in the
\MSbar{} scheme). The corresponding chirality-flipped operator in the JMS
basis is $\opL[]{dG}{12}^\dagger$. Class B operators contribute only via
mixing into $\OpL[(\prime)]{8g}$ operators, hence in \eqref{eq:master2}
\begin{align}
  P_b & = P_b^{\text{dipole}}. \qquad (\text{Class~B}) 
\end{align}
The class C and D operators have non-vanishing $K\to \pi\pi$ matrix elements
of their own, resulting in a contribution due to four-quark operators and
due to mixing a contribution from the dipole operator
\begin{align}
  P_b & = P_b^{\text{four-quark}} + P_b^{\text{dipole}} ,
  \qquad (\text{Class~C~and~D})
\end{align}
where the $P_b^{\text{four-quark}}$ are updated below at NLO in QCD.
On the other hand the $P_b^{\text{dipole}}$ can be found at LO for classes
C and D from tables 8 and 9 \cite{Aebischer:2018csl}, respectively.
As can be seen there, in practice $P_b^{\text{dipole}} \ll
P_b^{\text{four-quark}}$, such that within theory uncertainties one can
neglect for class C and D operators the $P_b^{\text{dipole}}$ due to mixing
into $\OpL{8g}$.

%--------+---------+---------+---------+---------+---------+---------+---------+
%
%
%
%--------+---------+---------+---------+---------+---------+---------+---------+
\section{Hadronic matrix elements}
\label{sec:3}

The calculation of the coefficients $P_b(\muEW)$ requires the knowledge of
the hadronic matrix elements
\begin{align}
  \langle \OpL{i} \rangle_I &
  \equiv \langle  (\pi\pi)_I | \OpL{i} | K \rangle (\mu)
&
  I & = 0,2 .
\end{align}
The number of $K\to \pi\pi$  matrix elements of four-quark operators in the
$\Nf = 3$ quark-flavour theory has been discussed in full generality in
\cite{Aebischer:2018csl}. Here we recall the most important points:
\begin{itemize}
\item
  The matrix elements of the chirality-flipped operators are related to the
  non-flipped ones due to parity conservation as $\langle \OpL[\prime]{i}
  \rangle_I = - \langle \OpL{i} \rangle_I$, thereby reducing the number of
  independent matrix elements by a factor of two.
\item
  We assume that operators with quark-flavour content $(\oL{s}d)(\oL{s}s)$
  are strongly suppressed and neglect them, see explanations in
  \cite{Aebischer:2018csl}. In practice this concerns only the BSM operators,
  because for matrix elements of SM operators the LQCD calculations include
  these contributions.
\item
  For the operators with quark-flavour content $(\oL{s}d)(\oL{u}u)$ and
  $(\oL{s}d)(\oL{d}d)$ there are $15_0$ and~$8_2$ matrix elements for
  $I = 0$ and $I = 2$, respectively, when using exact or nearly exact
  symmetries of parity and isospin. Out of these, $7_0$ and $3_2$ are
  known from the SM operators from LQCD calculations \cite{RBC:2020kdj,
  Blum:2015ywa}. The remaining $8_0$ and $2_2$ matrix elements of the BSM
  operators are available from the DQCD calculation \cite{Aebischer:2018rrz}.
\end{itemize}

As already stressed in the previous section it is useful to perform this
calculation in the BMU operator basis because in this basis the RG evolution
is available at NLO in QCD, contrary to the JMS basis.
The SM operators are part of the BMU basis and their matrix elements from
the LQCD calculations \cite{RBC:2020kdj} for $I = 0$ and \cite{Blum:2015ywa}
for $I = 2$ can be used after RG evolution from the scales $\mu_0 = 4.006\GeV$
and $\mu_2 = 3.0\GeV$, respectively, to the scale $\mu = 1.3\GeV$ that we will
use here. The results and details of the RG evolution are given in
\cite{Aebischer:2020jto}.\footnote{As explained in \cite{Aebischer:2020jto},
the perturbative RG flow contains isospin-breaking corrections from quark-charges
and leads to deviations from the isospin limits of the matrix elements after
evolution.}

The matrix elements of the BSM operators from the DQCD calculation
\cite{Aebischer:2018rrz} are given in the BMU basis, with the exception
of the SRR,$d$ sector shown in \refapp{app:BMU-SD}, at the scale $\muLow =
1.0\GeV$. The results after the RG evolution to $\mu = 1.3\GeV$ are provided
in \cite{Aebischer:2020jto}. In the case of the SRR,$d$ sector Fierz identities
have to be used in order to relate BMU operators to the ones used in
\cite{Aebischer:2018rrz}. To this end we will use tree-level relations and
thereby neglect possible effects of evanescent operators required
for a consistent treatment at NLO in QCD, simply because the present accuracy
of the matrix elements obtained in the DQCD approach is at best at the level
of $10\%$, such that effects at NLO in QCD can be neglected.
However, we kept in \cite{Aebischer:2021raf} the NLO corrections in
the evaluation of the RG evolution contained in $p_{ba}^{(I)}(\muEW, \muLow)$.
Indeed, anticipating more accurate future results from LQCD that should be
performed directly in the BMU basis, such effects can be included consistently
one day.

\begin{table}%[!htb]
\centering
\renewcommand{\arraystretch}{1.3} 
\begin{adjustbox}{width=0.95\textwidth}  
%\begin{tabular}{|lll|lll|}
\begin{tabular}{|l D{.}{.}{14} D{.}{.}{13} |l D{.}{.}{13} D{.}{.}{12}|}
\toprule
  Op. 
& \multicolumn{1}{c}{$\langle Q_i \rangle_0$}
& \multicolumn{1}{c|}{$\langle Q_i \rangle_2$}
& Op.
& \multicolumn{1}{c}{$\langle Q_i \rangle_0$}
& \multicolumn{1}{c|}{$\langle Q_i \rangle_2$} \\
\hline \hline 
$Q_1$    & -0.0163(43)(25)   &  0.0021(1)(1)    & $Q_{15}$ & -0.3708(92)(448) & -0.1138(32)(68) \\
$Q_2$    &  0.0217(32)(35)   &  0.0021(1)(1)    & $Q_{16}$ & -0.0562(92)(65)  & -0.0165(11)(5)  \\
$Q_3$    & -0.0187(143)(30)  &                  & $Q_{19}$ &  0.0150(30)      & -0.0030(6)      \\
$Q_4$    &  0.0232(130)(37)  &                  & $Q_{20}$ &  0.1410(280)     & -0.0500(100)    \\
$Q_9$    & -0.0152(43)(22)   &  0.0032(1)(2)    & $Q_{25}$ &  0.0880(180)     & -0.0310(60)     \\
$Q_{10}$ &  0.0230(45)(35)   &  0.0032(1)(2)    & $Q_{26}$ & -0.2960(729)     &  0.1080(400)    \\
$Q_{14}$ & -0.0025(149)(39)  & -0.0021(1)(1)    & $Q_{29}$ &  0.0050(10)      &  0.0030(6)      \\
$Q_5$    & -0.0302(132)(47)  &                  & $Q_{30}$ &  0.0440(90)      &  0.0310(60)     \\
$Q_6$    & -0.1610(115)(253) &                  & $Q_{31}$ & -0.3710(740)     & -0.2620(520)    \\
$Q_7$    &  0.0540(40)(85)   &  0.0247(17)(8)   & $Q_{32}$ & -0.2140(430)     & -0.1510(300)    \\
$Q_8$    &  0.3952(75)(622)  &  0.1708(47)(103) &          &                  &                 \\
\bottomrule
\end{tabular}

\end{adjustbox}
\renewcommand{\arraystretch}{1.0}
\caption{\label{tab:mebmu}
  The matrix elements ${\langle \OpL{i} \rangle}_{0,2}$ [$\geV^3$] for $\Nf=3$
  BMU basis at scale $\mu = 1.3\GeV$ are shown. The statistical and systematic
  uncertainties are given for results from lattice calculations
  $\OpL{1,\ldots,16}$, whereas only parametric uncertainties for results
  from DQCD calculations $\OpL{19, \ldots, 32}$.
}
\end{table}

Working then for $N_f=3$, but neglecting $(\bar s s)$ contributions, most
BSM operators in the BMU basis coincide with the ones in the basis of
\cite{Aebischer:2018rrz} and hence have the same matrix
elements.\footnote{Equal up to $P_L \leftrightarrow P_R$, which implies the
flip of sign in $K\to\pi\pi$ matrix elements, or differing by a sign due to
different definitions of tensor operators.} In consequence they can be directly
obtained from the DQCD results in \cite{Aebischer:2018rrz} at $\mu = 1.3\GeV$
with appropriate RG evolution. These BSM operators in the
BMU basis are
\begin{align}
  \OpL{i}, \qquad i = 14,15,16,\; 19,20,\; 25,26,\; 29,30,31,32\,.
\end{align}
The operators $\OpL{14, 15, 16}$ become linearly dependent on the SM operators
when neglecting the $(\bar s s)$ contributions and we can obtain their
matrix elements in this approximation as follows:
\begin{align}
  \langle \OpL{14} \rangle_I &
  \simeq \langle\OpL{3} - \OpL{1} \rangle_I \,,
&
  \langle \OpL{15} \rangle_I &
  \simeq \frac23 \langle \OpL{6} - \OpL{8} \rangle_I \,,
&
  \langle \OpL{16} \rangle_I &
  \simeq \frac23 \langle \OpL{5} - \OpL{7} \rangle_I \,,
\end{align}
from known LQCD results.\footnote{Based on the
assumption that the $(\oL{s}s)$ contribution is strongly suppressed.}
This leaves us with only 8 BSM operators $(Q_{19,20},Q_{25,26},Q_{29-32})$,
that can be mapped onto the SD basis in \cite{Aebischer:2018rrz}. Among these,
the only BSM operator in the BMU basis for which a Fierz transformation is
required is $Q_{26}$, see \eqref{eq:BMU->SD}. The complete basis transformation
of the BMU operators to the SD basis of \cite{Aebischer:2018rrz} is given in
\refapp{app:BMU-SD} and allows to determine the $I = 0, 2$ matrix
elements of these operators at the input scale $\muLow = 1.0\GeV$.

We collect the values of hadronic matrix elements in \reftab{tab:mebmu}.
We notice in particular that the matrix elements of the operators $\OpL{i}$
with $i=6, 8, 15, 20,26,31,32$ have the largest matrix elements. These are
the VLR, scalar and tensor operators. In comparing the values of the matrix
elements of the SM operators with those present in the literature one
should take into account that our SM operators are defined with $P_{L,R}$
instead of traditional $V\mp A$, which brings in a factor of four:
$\langle \OpL[\text{our}]{i} \rangle = \langle \OpL[\text{trad}]{i}
\rangle/4$. The chromomagnetic dipole operator has the nonvanishing
matrix element ${\langle \OpL[\prime]{8g} \rangle}_{0} = +0.013(4)\GeV^3$ 
\cite{Constantinou:2017sgv, Buras:2018evv}.

%--------+---------+---------+---------+---------+---------+---------+---------+
%
%
%
%--------+---------+---------+---------+---------+---------+---------+---------+
\section[Results for $P_b(\muEW)$ in the JMS basis]
{\bm Results for $P_b(\muEW)$ in the JMS basis}
\label{sec:4}

\begin{table}[!t]
\centering
\renewcommand{\arraystretch}{1.3}
\begin{tabular}{|lrr|lrr|} 
\toprule
& \multicolumn{2}{c|}{$P_{b}(\muEW)$}
& 
& \multicolumn{2}{c|}{$P_{b}(\muEW)$}
\\
  $C_b (\muEW)$ & \multicolumn{1}{c}{NLO} & \multicolumn{1}{c|}{LO}
& $C_b (\muEW)$ & \multicolumn{1}{c}{NLO} & \multicolumn{1}{c|}{LO}
\\
\hline\hline
  $[C_{ud}^{V1,LL}]_{1121}$   &   -3.47 &   -3.34 & $[C_{ud}^{V8,LL}]_{1121}$   &    0.77 &    0.63 \\
  $[C_{ud}^{V1,LL}]_{2221}$   &   -0.21 &   -0.13 & $[C_{ud}^{V8,LL}]_{2221}$   &    0.39 &    0.28 \\
  $[C_{dd}^{VLL}]_{2133}$     &   -0.12 &   -0.08 & $[C_{dd}^{VLL}]_{2331}$     &    0.53 &    0.46 \\
  $[C_{dd}^{VLL}]_{2111}$     &    2.26 &    2.15 & $[C_{dd}^{VLL}]_{2122}$     &    0.46 &    0.36 \\
  $[C_{du}^{V1,LR}]_{2111}$   & -142.16 & -108.08 & $[C_{du}^{V8,LR}]_{2111}$   & -222.69 & -174.02 \\
  $[C_{du}^{V1,LR}]_{2122}$   &    0.27 &    0.10 & $[C_{du}^{V8,LR}]_{2122}$   &    0.70 &    0.33 \\
  $[C_{dd}^{V1,LR}]_{2133}$   &    0.10 &    0.04 & $[C_{dd}^{V8,LR}]_{2133}$   &    0.39 &    0.20 \\
  $[C_{dd}^{V1,LR}]_{2111}$   &  134.07 &  101.82 & $[C_{dd}^{V8,LR}]_{2111}$   &  210.94 &  164.71 \\
  $[C_{dd}^{V1,LR}]_{2122}$   &    0.18 &    0.10 & $[C_{dd}^{V8,LR}]_{2122}$   &    0.53 &    0.33 \\
  $[C_{uddu}^{V1,LR}]_{1211}$ &  139.10 &  103.83 & $[C_{uddu}^{V8,LR}]_{1211}$ &  233.47 &  183.69 \\
  $[C_{dd}^{S1,RR}]_{2111}$   &   92.02 &   78.90 & $[C_{dd}^{S8,RR}]_{2111}$   &  -28.05 &  -21.69 \\
  $[C_{ud}^{S1,RR}]_{1121}$   &  -86.98 &  -66.64 & $[C_{ud}^{S8,RR}]_{1121}$   &   18.09 &   17.26 \\
  $[C_{uddu}^{S1,RR}]_{1121}$ &  -56.42 &  -67.69 & $[C_{uddu}^{S8,RR}]_{1121}$ &   22.65 &   16.99 \\
%%%%%%%%%%%%%%%%%%%%%%%%%%%%%%%%%%%%%%%%%                                                 
\hline\hline
  $[C^{S1,RR}_{dd}]_{2122}$   &         &   -0.05  & $[C^{S8,RR}_{dd}]_{2122}$   &         &   0.01 \\
  $[C^{S1,RR}_{ud}]_{2221}$   &         &    0.26  & $[C^{S8,RR}_{ud}]_{2221}$   &         &   0.07 \\ 
  $[C^{S1,RR}_{uddu}]_{2221}$ &         &   -0.85  & $[C^{S8,RR}_{uddu}]_{2221}$ &         &   0.07 \\
  $[C^{S1,RR}_{dd}]_{2133}$   &         &    0.35  & $[C^{S8,RR}_{dd}]_{2133}$   &         &   0.11 \\
  $[C^{S1,RR}_{dd}]_{2331}$   &         &   -1.85  & $[C^{S8,RR}_{dd}]_{2331}$   &         &   0.21 \\
 \bottomrule 
\end{tabular}

\renewcommand{\arraystretch}{1.0}
\caption{\label{Pbtab}
  The numerical values for the quantities $P_b(\muEW)$ that enter in the
  master formula \eqref{eq:master} are shown for all JMS WCs contributing
  to $\epe$ for $\muEW = 160 \GeV$. The upper part is due to operators
  with non-vanishing $K\to\pi\pi$ matrix elements at NLO in QCD with
  LO results for comparison. The lower part is due to operators that
  contribute via mixing into $\OpL{8g}$ at LO in QCD.
}
\end{table}

In \reftab{Pbtab} we give the values of $P_b(\muEW)$ entering in the master formula
\eqref{eq:master} at the electroweak scale. These values are self-explanatory
and confirm what one would expect on the basis of the values of hadronic
matrix elements and the size of the anomalous dimensions: the coefficients
$P_b$ that multiply the WCs of left-right and scalar operators are typically
by one order of magnitude and in a few cases by two orders of magnitude
larger than of vector left-left and vector right-right operators.
However, the importance of a given contribution to $\epe$ will eventually
depend on the size of WCs at the electroweak scale and this will depend on
NP scenario considered. We recall that there are four out of the 40
four-quark operators that do not contribute to $\epe$ \cite{Aebischer:2018quc,
Aebischer:2018csl}, neither directly via a non-vanishing $K\to\pi\pi$ matrix
element nor via RG mixing. They correspond to the operators
$\opL[V1,LR]{uddu}{2212}$, $\opL[V8,LR]{uddu}{2212}$, $\opL[V1,LR]{dd}{2331}$
and $\opL[V8,LR]{dd}{2331}$ in the JMS basis.

We departed from the previous treatment of certain BSM matrix elements
in the master formula \cite{Aebischer:2018quc, Aebischer:2018csl,
Aebischer:2020jto}, by using here directly DQCD results \cite{Aebischer:2018rrz}
without employing isospin relations for $I=2$ matrix elements. Using instead
the previous treatment, following entries change
\begin{align}
  P_{[C_{uddu}^{V1,LR}]_{1211}} & = 131.68 , 
& P_{[C_{uddu}^{V8,LR}]_{1211}} & = 263.26 , \notag 
\\
  P_{[C_{ud}^{S1,RR}]_{1121}}   & = -86.80 ,
& P_{[C_{ud}^{S8,RR}]_{1121}}   & =  19.54 ,
\\
  P_{[C_{uddu}^{S1,RR}]_{1121}} & = -58.11 ,
& P_{[C_{uddu}^{S8,RR}]_{1121}} & =  24.21 . \notag
\end{align}
The changes are in the ballpark of $10\%$, which is in general smaller
compared to the uncertainties of about $20\%$ of the matrix elements in
the DQCD approach.

It is however clear that this table will be modified with time
when LQCD will improve the values of the SM hadronic matrix elements
and calculate the corresponding matrix elements for BSM operators.
Whether also in this case there will be consensus between LQCD and DQCD
on the  $I=2$ matrix elements remains to be seen. Such contributions are
expected to dominate NP contributions to $\epe$ because of  the
factor $\re A_2 / \re A_0  \sim 1/22$. However, to obtain the full picture
some consensus has to be reached on $I=0$ matrix elements, in particular
within the SM because this determines the room left for NP contributions.

In using the master formula \eqref{eq:master} it is crucial to take into
account the definition of Wilson coefficients as given in \eqref{eq:abbs}.
That is they enter the Hamiltonian with the minus sign and have
the dimension $\text{energy}^{-2}$. While it is convenient to use the
units $1/\TeV^2$, it is not essential because at the end the final results
for $P_b$ are dimensionless.

The theoretical uncertainties of the coefficients $P_b$ are determined firstly
by the uncertainties of the hadronic matrix elements 
${\langle \OpL{i} \rangle}_{0,2}$, listed in \reftab{tab:mebmu}. They enter
the $P_b$ linearly and result in $10\% - 50\%$ relative uncertainty,
depending on the $P_b$ \cite{Aebischer:2018quc}.
Secondly, scheme dependences due to higher order corrections in the
RG evolution are reduced here due to the inclusion of the NLO QCD
corrections in the RG evolution and the threshold corrections. 
As discussed in \cite{Aebischer:2021raf}, such effects are considerably
reduced now. The comparison of the LO results that do not include NLO QCD
corrections in the RG evolution of the Wilson coefficients in \reftab{Pbtab}
shows sizeable effects. They are up to $60\%$ for left-right vector operators,
but also for left-left vector operators these corrections can become larger
than $30\%$, whereas for the scalar right-right operators they are up to
$25\%$. The NLO QCD effects of the RG evolution are sizeable, but unfortunately
these improvements can not be fully appreciated in view of the remaining
large hadronic uncertainties of the matrix elements.

%--------+---------+---------+---------+---------+---------+---------+---------+
%
%
%
%--------+---------+---------+---------+---------+---------+---------+---------+
\section{Conclusions}
\label{sec:5}

The main result of our paper is the master formula \eqref{eq:master} with
the numerical values of the coefficients $P_b(\muEW)$ given in \reftab{Pbtab}.
The advantage of this formula over the one presented in \cite{Aebischer:2018quc}
is the inclusion of {next-to-leading order (NLO) QCD corrections in the RG evolution
in the Weak Effective Theory (WET) from the hadronic scale $\muLow$ to the
electroweak scale $\muEW$ and} the use of the JMS basis \cite{Jenkins:2017jig}.
NLO QCD evolution allows a correct matching of WCs to the matrix elements
calculated by lattice QCD (LQCD) or other non-perturbative methods sensitive
to renormalization scheme dependences. The use of the JMS basis on the other
hand allows to generalize this formula to the {Standard Model Effective
Field Theory (SMEFT)} one because in this basis the tree-level matching of
SMEFT on to WET \cite{Jenkins:2017jig} and the one-loop matching
\cite{Dekens:2019ept, Aebischer:2015fzz} are known.

Our master formula can already be used for approximate  estimates of BSM
contributions to $\epe$ using the JMS basis {in the SMEFT}. It constitutes
an important step towards a complete NLO QCD analysis of the ratio $\epe$
within the SMEFT. However, in order to finish this very ambitious project
the following steps have still to be performed:
\begin{itemize}
\item
  The $K\to\pi\pi$ matrix elements of all operators of the {BMU} basis, in
  particular of the BSM ones, should be calculated by LQCD and also the ones
  already found using DQCD \cite{Aebischer:2018rrz} could be in principle
  improved.
\item The contributions due to the chromomagnetic dipole operator
  and mixing of four-quark operators into the former, including threshold
  corrections, need to be extended to the NLO level.
\item
  The tree-level and one-loop matchings of SMEFT on WET that are already known
  \cite{Jenkins:2017jig, Dekens:2019ept, Aebischer:2015fzz} should be included.
  We have not done it because consistently also the next step should be made
  simultaneously.
\item
  The RG QCD evolution within the SMEFT up to the NP scale should be
  generalized to the NLO level, i.e. using two-loop anomalous dimensions.
\end{itemize}
Only then one will be able to cancel all unphysical renormalization scheme
dependences that we discussed in \cite{Aebischer:2021raf} at length.

While the second step can be performed already now and the third one could
be accomplished relatively soon, the first one is most difficult and could
still  take several years of intensive calculation by several LQCD
collaborations. We are looking forward to the improvements of our formula
and hope to report on the progress in a not too distant future.

%--------+---------+---------+---------+---------+---------+---------+---------+
\section*{Acknowledgements}

J. A. acknowledges financial support from the Swiss National Science Foundation
(Project No. P400P2\_183838).
The work of C.B. is supported by DFG under grant BO-4535/1-1.
A.J.B acknowledges financial support from the Excellence Cluster ORIGINS,
funded by the Deutsche Forschungsgemeinschaft (DFG, German Research Foundation)
under Germany´s Excellence Strategy – EXC-2094 – 390783311.
J.K. is financially supported by the Alexander von Humboldt Foundation's
postdoctoral research fellowship.

%--------+---------+---------+---------+---------+---------+---------+---------+
%
%
%
%--------+---------+---------+---------+---------+---------+---------+---------+

\appendix

%--------+---------+---------+---------+---------+---------+---------+---------+
%
%
%
%--------+---------+---------+---------+---------+---------+---------+---------+
\section{JMS operator basis}
\label{app:def-JMS}

The JMS basis \cite{Jenkins:2017jig} categorizes dimension-six four-quark
operators with four different chirality structures $(\oL{L}L)(\oL{L}L)$,
$(\oL{R}R)(\oL{R}R)$, $(\oL{L}L)(\oL{R}R)$ and $(\oL{L}R)(\oL{L}R)$.
The non-leptonic operators that contain at least one pair of down-type quarks
are given as
\begin{align}
  \notag
  & (\oL{L}L)(\oL{L}L) &
\\ \notag
  \opL[VLL]{dd}{prst} &
  = (\bar{d}_L^p \gamma_\mu d_L^r) (\bar{d}_L^s \gamma^\mu d_L^t) , &
\\
  \opL[V1,LL]{ud}{prst} &
  = (\bar{u}_L^p \gamma_\mu u_L^r) (\bar{d}_L^s \gamma^\mu d_L^t) , &
  \opL[V8,LL]{ud}{prst} &
  = (\bar{u}_L^p \gamma_\mu T^A u_L^r) (\bar{d}_L^s \gamma^\mu T^A d_L^t) ,
\\[0.2cm]
  \notag
  & (\oL{R}R)(\oL{R}R) &
\\
  \notag
  \opL[VRR]{dd}{prst} &
  = (\bar{d}_R^p \gamma_\mu d_R^r) (\bar{d}_R^s \gamma^\mu d_R^t) , &
\\
  \opL[V1,RR]{ud}{prst} &
  = (\bar{u}_R^p \gamma_\mu u_R^r) (\bar{d}_R^s \gamma^\mu d_R^t) , &
  \opL[V8,RR]{ud}{prst} &
  = (\bar{u}_R^p \gamma_\mu T^A u_R^r) (\bar{d}_R^s \gamma^\mu T^A d_R^t) ,
\\[0.2cm]
  \notag
  & (\oL{L}L)(\oL{R}R) &
\\
  \notag
  \opL[V1,LR]{dd}{prst} &
  = (\bar{d}_L^p \gamma_\mu d_L^r) (\bar{d}_R^s \gamma^\mu d_R^t) , &
  \opL[V8,LR]{dd}{prst} &
  = (\bar{d}_L^p \gamma_\mu T^A d_L^r) (\bar{d}_R^s \gamma^\mu T^A d_R^t) ,
\\
  \notag
  \opL[V1,LR]{ud}{prst} &
  = (\bar{u}_L^p \gamma_\mu u_L^r) (\bar{d}_R^s \gamma^\mu d_R^t) , &
  \opL[V8,LR]{ud}{prst} &
  = (\bar{u}_L^p \gamma_\mu T^A u_L^r) (\bar{d}_R^s \gamma^\mu T^A d_R^t) ,
\\
  \notag
  \opL[V1,LR]{du}{prst} &
  = (\bar{d}_L^p \gamma_\mu d_L^r) (\bar{u}_R^s \gamma^\mu u_R^t) , &
  \opL[V8,LR]{du}{prst} &
  = (\bar{d}_L^p \gamma_\mu T^A d_L^r) (\bar{u}_R^s \gamma^\mu T^A u_R^t) ,
\\
  \opL[V1,LR]{uddu}{prst} &
  = (\bar{u}_L^p \gamma_\mu d_L^r) (\bar{d}_R^s \gamma^\mu u_R^t) , &
  \opL[V8,LR]{uddu}{prst} &
  = (\bar{u}_L^p \gamma_\mu T^A d_L^r) (\bar{d}_R^s \gamma^\mu T^A u_R^t) ,
\\[0.2cm]
  \notag
  & (\oL{L}R)(\oL{L}R) &
\\
  \notag
  \opL[S1,RR]{dd}{prst} &
  = (\bar{d}_L^p d_R^r) (\bar{d}_L^s d_R^t) , &
  \opL[S8,RR]{dd}{prst} &
  = (\bar{d}_L^p T^A d_R^r) (\bar{d}_L^s T^A d_R^t) ,
\\
  \notag
  \opL[S1,RR]{ud}{prst} &
  = (\bar{u}_L^p u_R^r) (\bar{d}_L^s d_R^t) , &
  \opL[S8,RR]{ud}{prst} &
  = (\bar{u}_L^p T^A u_R^r) (\bar{d}_L^s T^A d_R^t) ,
\\
  \opL[S1,RR]{uddu}{prst} &
  = (\bar{u}_L^p d_R^r) (\bar{d}_L^s u_R^t) , &
  \opL[S8,RR]{uddu}{prst} &
  = (\bar{u}_L^p T^A d_R^r) (\bar{d}_L^s T^A u_R^t) ,
\end{align}
Some of the listed operators are not equal to their hermitian conjugates with
permuted generation indices. Although we do not list them explicitly here,
they have to be treated as independent operators. Note that also
$\OpL[V1,LR]{uddu}$ and $\OpL[V8,LR]{uddu}$ have Hermitian conjugates. The
same holds for the operators $(\oL{L}R)(\oL{L}R)$. This choice of basis
eliminates all operators with Dirac structures~$\sigma^{\mu\nu}$.
The colour structure is expressed in terms of the generators $T^A$ of
$\SUthreeC$. Above $p,r,s,t$ denote quark flavour indices in the
mass-eigenstate basis. These operators include both the ``non-flipped''
and the chirality-flipped ones.

%--------+---------+---------+---------+---------+---------+---------+---------+
%
%
%
%--------+---------+---------+---------+---------+---------+---------+---------+
\section[BMU operator basis for $K\to\pi\pi$]
{\bm BMU operator basis for $K\to\pi\pi$}
\label{app:def-BMU}

The BMU basis \cite{Buras:2000if} contains several sectors of operators,
depending on their quark-flavour content and Lorentz structures. They
form separate blocks under RG evolution, thereby minimizing the operator
mixing. General description of this basis valid also for the $B$-meson system
is discussed in detail in the Appendix A.2 in \cite{Aebischer:2021raf}. Here
we specify this basis to the case relevant for $K\to\pi\pi$ decays. This
amounts to setting $d_j=s$ and $d_i=d$ in the expressions listed in
\cite{Aebischer:2021raf}. Moreover in some operators $d_k=b$.

The {\bf SM basis} consists of two
current-current-type operators
\begin{equation}
\begin{aligned}
  \OpL{1} &
  = \OpL[\text{VLL},u]{1}
  = (\bar s^\alpha \gamma_\mu P_L u^\beta)
    (\bar u^\beta    \gamma^\mu P_L d^\alpha) ,
\\
  \OpL{2} &
  = \OpL[\text{VLL},u]{2}
  = (\bar s^\alpha \gamma_\mu P_L u^\alpha)
    (\bar u^\beta    \gamma^\mu P_L d^\beta) ,
\end{aligned}
\end{equation}
where the naming used in \cite{Buras:2000if} is given as well. The $\SUthreeC$
colour indices are denoted by $\alpha$ and $\beta$. The choice of $u_k = u$
implies that the corresponding operators for $u_k = c$ are eliminated in
favour of QCD- and QED-penguin operators, see below, which is most suited
for Kaon physics.

There are eight penguin-type operators
with left-handed $(\bar s \gamma_\mu P_L d)$ vector currents: the {\bf\bm P$_\text{QCD}$}
\begin{equation}
  \label{eq:QCD-peng-op}
\begin{aligned}
  \OpL{3} & = (\bar s^\alpha \gamma_\mu P_L d^\alpha)
    \!\sum_{q}(\bar q^\beta    \gamma^\mu P_L \, q^\beta) , \qquad
&%\\
  \OpL{4} & = (\bar s^\alpha \gamma_\mu P_L d^\beta)
  \!\sum_{q}(\bar q^\beta    \gamma^\mu P_L \, q^\alpha) ,
\\
  \OpL{5} & = (\bar s^\alpha \gamma_\mu P_L d^\alpha)
    \!\sum_{q}(\bar q^\beta    \gamma^\mu P_R \, q^\beta) ,
&%\\
  \OpL{6} & = (\bar s^\alpha \gamma_\mu P_L d^\beta)
    \!\sum_{q}(\bar q^\beta    \gamma^\mu P_R \, q^\alpha) ,
\end{aligned}
\end{equation}
and {\bf\bm P$_\text{QED}$}
\begin{equation}
  \label{eq:QED-peng-op}
\begin{aligned}
  \OpL{7} & = \frac{3}{2}\,(\bar s^\alpha \gamma_\mu P_L d^\alpha)
      \!\sum_{q} \! Q_q \, (\bar q^\beta \gamma^\mu P_R \, q^\beta) ,  \qquad
&%\\
  \OpL{8} & = \frac{3}{2}\,(\bar s^\alpha \gamma_\mu P_L d^\beta)
      \!\sum_{q} \! Q_q \, (\bar q^\beta \gamma^\mu P_R \, q^\alpha) ,
\\
  \OpL{9} & = \frac{3}{2}\,(\bar s^\alpha \gamma_\mu P_L d^\alpha)
      \!\sum_{q} \! Q_q \, (\bar q^\beta \gamma^\mu P_L \, q^\beta) ,
&%\\
 \OpL{10} & =\frac{3}{2}\,(\bar s^\alpha \gamma_\mu P_L d^\beta)
      \!\sum_{q} \! Q_q \, (\bar q^\beta \gamma^\mu P_L \, q^\alpha) .
\end{aligned}
\end{equation}
The sum over $q = u,d,s,c,b$ includes all five active quark flavours
and $Q_q$ their electromagnetic charges: $Q_u = +2/3$ and $Q_d = -1/3$.
Their chirality-flipped counterparts are not present in the SM.

Then the SM basis contains ten operators only because
the chirality-flipped operators are absent. No Wilson coefficients
are generated for them due to the left-handed nature of electroweak interactions
of the SM.

{\bf Beyond the SM}, we have first operators with vectorial Lorentz structure
\cite{Buras:2000if}
\begin{equation}
\begin{aligned}
  \OpL{11} = \; \OpL[\text{VLL},d+s]{1} &
  = (\bar s^\alpha \gamma_\mu P_L d^\alpha) \, \big[
    (\bar d^\beta \gamma^\mu P_L d^\beta)
  + (\bar s^\beta \gamma^\mu P_L s^\beta)\big] ,
\\
  \OpL{12} = \; \OpL[\text{VLR},d+s]{1} &
  = (\bar s^\alpha \gamma_\mu P_L d^\beta) \, \big[
    (\bar d^\beta \gamma^\mu P_R d^\alpha)
  + (\bar s^\beta \gamma^\mu P_R s^\alpha)\big] ,
\\
  \OpL{13} = \; \OpL[\text{VLR},d+s]{2} &
  = (\bar s^\alpha \gamma_\mu P_L d^\alpha) \, \big[
    (\bar d^\beta \gamma^\mu P_R d^\beta)
  + (\bar s^\beta \gamma^\mu P_R s^\beta)\big] ,
\end{aligned}
\end{equation}
  \begin{equation}
\begin{aligned}
  \OpL{14} = \; \OpL[\text{VLL},d-s]{1} &
  = (\bar s^\alpha \gamma_\mu P_L d^\alpha) \,\big[
    (\bar d^\beta \gamma^\mu P_L d^\beta)
  - (\bar s^\beta \gamma^\mu P_L s^\beta)\big] ,
\\
  \OpL{15} = \; \OpL[\text{VLR},d-s]{1} &
  = (\bar s^\alpha \gamma_\mu P_L d^\beta) \, \big[
    (\bar d^\beta \gamma^\mu P_R d^\alpha)
  - (\bar s^\beta \gamma^\mu P_R s^\alpha)\big] ,
\\
  \OpL{16} = \; \OpL[\text{VLR},d-s]{2} &
  = (\bar s^\alpha \gamma_\mu P_L d^\alpha) \, \big[
    (\bar d^\beta \gamma^\mu P_R d^\beta)
  - (\bar s^\beta \gamma^\mu P_R s^\beta)\big] .
\end{aligned}
  \end{equation}
Only $d+s$ operators mix into QCD- and QED-penguin operators $\OpL{3,\ldots 10}$,
whereas $d-s$ ones do not. Analogous operators for $d \to s \, u_k \bar u_k$
are
\begin{equation}
\begin{aligned}
  \OpL{17} = \; \OpL[\text{VLR}, u-c]{1} &
  = (\bar s^\alpha \gamma_\mu P_L d^\beta) \, \big[
    (\bar u^\beta \gamma^\mu P_R \, u^\alpha)
  - (\bar c^\beta \gamma^\mu P_R \, c^\alpha) \big],
\\
  \OpL{18} = \; \OpL[\text{VLR}, u-c]{2} &
  = (\bar s^\alpha \gamma_\mu P_L d^\alpha) \, \big[
    (\bar u^\beta \gamma^\mu P_R \, u^\beta)
  - (\bar c^\beta \gamma^\mu P_R \, c^\beta) \big],
\end{aligned}
\end{equation}
which do not mix into QCD- and QED-penguin operators $\OpL{3,\ldots 10}$, but
give threshold corrections when decoupling the $c$ quark.

The operators of the {\bf SRL sector} for $d \to s \, Q \oL{Q}$ with
$Q = u,c,b$ are given by
\cite{Buras:2000if}
\begin{align}
  \OpL[\text{SRL}, Q]{1} &
  = (\bar s^\alpha P_R d^\beta) \, (\bar Q^\beta P_L \, Q^\alpha) ,
&
  \OpL[\text{SRL}, Q]{2} &
  = (\bar s^\alpha P_R d^\alpha) \, (\bar Q^\beta P_L \, Q^\beta) ,
\end{align}
which are numbered as
\begin{equation}
\begin{aligned}
  (\OpL{19},\, \OpL{20}) & = (\OpL[\text{SRL}, u]{1},\, \OpL[\text{SRL}, u]{2}) ,
  \qquad \qquad
&
  (\OpL{23},\, \OpL{24}) & = (\OpL[\text{SRL}, b]{1},\, \OpL[\text{SRL}, b]{2}) .
\\
  (\OpL{21},\, \OpL{22}) & = (\OpL[\text{SRL}, c]{1},\, \OpL[\text{SRL}, c]{2}) ,
\end{aligned}
\end{equation}
The scalar operators of the {\bf SRR sectors} for $d \to s \, d \oL{d}$
and $d \to s \, s \oL{s}$ have been chosen in \cite{Buras:2000if}
as colour-singlet operators, see $\DF = 2$ sector. Due to Fierz symmetries
there are only two operators per sector SRR,$d$
\begin{align}
  \OpL{25} = \OpL[\text{SRR}, d]{1} &
  = (\bar s^\alpha P_R \, d^\alpha) \, (\bar d^\beta P_R \, d^\beta) ,
&
  \OpL{26} = \OpL[\text{SRR}, d]{2} &
  = (\bar s^\alpha \sigma_{\mu\nu} P_R \, d^\alpha) \,
    (\bar d^\beta  \sigma^{\mu\nu} P_R \, d^\beta) ,
\intertext{and SRR$,s$}
  \OpL{27} = \OpL[\text{SRR}, s]{1} &
  = (\bar s^\alpha P_R \, d^\alpha) \, (\bar s^\beta P_R \, s^\beta) ,
&
  \OpL{28} = \OpL[\text{SRR}, s]{2} &
  = (\bar s^\alpha \sigma_{\mu\nu} P_R \, d^\alpha) \,
    (\bar s^\beta  \sigma^{\mu\nu} P_R \, s^\beta) .
\end{align}
The case of $d \to s \, Q \oL{Q}$ with $Q = u,c,b$
comprises four operators per sector
\begin{equation}
  \label{eq:def-BMU-SRR-ops}
\begin{aligned}
  \OpL[\text{SRR}, Q]{1} &
  = (\bar s^\alpha P_R \, d^\beta) \, (\bar Q^\beta P_R \, Q^\alpha) , \qquad
&
  \OpL[\text{SRR}, Q]{3} &
  = (\bar s^\alpha \sigma_{\mu\nu} P_R \, d^\beta) \,
    (\bar Q^\beta    \sigma^{\mu\nu} P_R \, Q^\alpha) ,
\\
  \OpL[\text{SRR}, Q]{2} &
  = (\bar s^\alpha P_R \, d^\alpha) \, (\bar Q^\beta P_R \, Q^\beta) ,
&\OpL[\text{SRR}, Q]{4} &
  = (\bar s^\alpha \sigma_{\mu\nu} P_R \, d^\alpha) \,
    (\bar Q^\beta    \sigma^{\mu\nu} P_R \, Q^\beta) .
\end{aligned}
\end{equation}
They are numbered consecutively
\begin{equation}
\begin{aligned}
  (\OpL{29},\, \OpL{30},\, \OpL{31},\, \OpL{32}) &
  = (\OpL[\text{SRR}, u]{1},\, \OpL[\text{SRR}, u]{2},\,
     \OpL[\text{SRR}, u]{3},\, \OpL[\text{SRR}, u]{4}) ,
\\
  (\OpL{33},\, \OpL{34},\, \OpL{35},\, \OpL{36}) &
  = (\OpL[\text{SRR}, c]{1},\,\, \OpL[\text{SRR}, c]{2},\,\,
     \OpL[\text{SRR}, c]{3},\,\, \OpL[\text{SRR}, c]{4}) ,
\\
  (\OpL{37},\, \OpL{38},\, \OpL{39},\, \OpL{40}) &
  = (\OpL[\text{SRR}, b]{1},\, \OpL[\text{SRR}, b]{2},\,
     \OpL[\text{SRR}, b]{3},\, \OpL[\text{SRR}, b]{4}) .
\end{aligned}
\end{equation}

The definition of chirality-flipped operators is not explicitly
shown here. The numbering of these operators is given as
\begin{align}
  \OpL{40 + i} & = \OpL{i}[P_L \leftrightarrow P_R] ,
\end{align}
i.e. they are found by interchange of $P_L \leftrightarrow P_R$ from the
``non-flipped'' operators.

%--------+---------+---------+---------+---------+---------+---------+---------+
%
%
%
%--------+---------+---------+---------+---------+---------+---------+---------+
\section{Basis transformation: BMU to SD}
\label{app:BMU-SD}

In this appendix we report the basis change from the BMU basis used in the
RG evolution to the SD basis defined in \cite{Aebischer:2018rrz}, in
which the matrix elements for the BSM operators are given. Both
bases almost coincide, except for a few exceptions. The relations
for the BSM operators $\OpL{i}$ in the BMU basis and $O_i$ in the SD
basis, respectively, read:
\begin{equation}
  \label{eq:BMU->SD}
\begin{aligned}
  \OpL{14} &= O^{\text{VLL},d-s}_{1}\,,
& \qquad\quad &
\\[0.3cm]
  \OpL{15} &= O^{\text{VLR},d-s}_{1}\,,
&
  \OpL{16} &= O^{\text{VLR},d-s}_{2}\,,
\\[0.1cm]
  \OpL{19} &= (O^{\text{SLR},u}_{1})^\prime = O^{\text{SRL},u}_{1}\,,
&
  \OpL{20} &= (O^{\text{SLR},u}_{2})^\prime = O^{\text{SRL},u}_{2}\,,
\\[0.3cm]
  \OpL{29} &= (O^{\text{SLL},u}_{1})^\prime =  O^{\text{SRR},u}_{1}\,,
&
  \OpL{31} &=-(O^{\text{SLL},u}_{3})^\prime = -O^{\text{SRR},u}_{3}\,,
\\
  \OpL{30} &= (O^{\text{SLL},u}_{2})^\prime =  O^{\text{SRR},u}_{2}\,,
&
  \OpL{32} &=-(O^{\text{SLL},u}_{4})^\prime = -O^{\text{SRR},u}_{4}\,,
\\[0.3cm]
  \OpL{25} &= (O^{\text{SLL},d}_{2})^\prime = O^{\text{SRR},d}_{2}\,,
&
  \OpL{26} &= -8(O^{\text{SLL},d}_{1})^\prime - 4 (O^{\text{SLL},d}_{2})^\prime
            = -8 O^{\text{SRR},d}_{1} - 4 O^{\text{SRR},d}_{2}\,,
\\[0.3cm]
  \OpL{27} &= (O^{\text{SLL},s}_{2})^\prime = O^{\text{SRR},s}_{2}\,,
&
  \OpL{28} &= -8(O^{\text{SLL},s}_{1})^\prime - 4 (O^{\text{SLL},s}_{2})^\prime
            = -8 O^{\text{SRR},s}_{1} - 4 O^{\text{SRR},s}_{2}\,.
\end{aligned}
\end{equation}

%--------+---------+---------+---------+---------+---------+---------+---------+
%
%
%
%--------+---------+---------+---------+---------+---------+---------+---------+
\section{Basis transformation: BMU to JMS}
\label{app:SD2-BMU}

Here we report the basis transformation from the basis used in
\cite{Aebischer:2018quc, Aebischer:2018csl, Aebischer:2020mkv}, which
coincides with the BMU basis, to the JMS basis. This basis change allows
us to use the results obtained in \cite{Aebischer:2020mkv} for the operators
in class B \cite{Aebischer:2018quc}, which consists of operators that only
contribute to $\epe$ via RG mixing into the dipole operators
$\OpL[(\prime)]{8g}$. The mixing of these operators was computed using
\texttt{wilson} \cite{Aebischer:2018bkb}, which includes the one-loop
QCD and QED \cite{Aebischer:2017gaw, Jenkins:2017dyc} running below the
EW scale. In the notation of \cite{Aebischer:2018quc} we find the following
relations:
\begin{align}
  O^s_{SRR} & = \OpL{27}
  = \opL[S1,RR]{dd}{2122} \,,
\\
  O^s_{TRR} & = \OpL{28}
  = -\frac{20}{3}\opL[S1,RR]{dd}{2122} - 16 \opL[S8,RR]{dd}{2122}
\\
  O^c_{SRR}  & = \OpL{34}
  = \opL[S1,RR]{ud}{2221}\,,
\\
  O^c_{TRR}  & = \OpL{36}
  = -\frac{8}{3}\opL[S1,RR]{uddu}{2122}
    -4\opL[S1,RR]{ud}{2221} - 16\opL[S8,RR]{uddu}{2122} \,,
\\
  \wT O^c_{SRR} & = \OpL{33}
  = \frac{1}{3}\opL[S1,RR]{ud}{2221} + 2\opL[S8,RR]{ud}{2221} \,,
\\
  \wT O^c_{TRR} & = \OpL{35}
  =-8\opL[S1,RR]{uddu}{2122}
  -\frac{4}{3}\opL[S1,RR]{ud}{2221} - 8\opL[S8,RR]{ud}{2221}\,,
\\
  O^b_{SRR} & = \OpL{38}
  = \opL[S1,RR]{dd}{2133} \,,
\\
  O^b_{TRR} & = \OpL{40}
  = -\frac{8}{3}\opL[S1,RR]{dd}{2331}
  -4\opL[S1,RR]{dd}{2133} - 16\opL[S8,RR]{dd}{2331} \,,
\\
  \wT O^b_{SRR}  &= \OpL{37}
  =\frac{1}{3}\opL[S1,RR]{dd}{2133} + 2\opL[S8,RR]{dd}{2133} \,,
\\
  \wT O^b_{TRR}  &= \OpL{39}
  = -8\opL[S1,RR]{dd}{2331}
  - \frac{4}{3}\opL[S1,RR]{dd}{2133} - 8\opL[S8,RR]{dd}{2133} \,.
\end{align}

% For the Wilson coefficients one finds:
% \begin{equation}
%   \vec{C}_\text{BMU} = \hat M \, \vec{C}_\text{JMS}\,,
% \end{equation}
% 
% with
% \begin{align}
%   \hat M & = \begin{pmatrix}
%   \hat A & 0 & 0\\
%   0 & \hat B & 0\\
%   0 & 0 & \hat B
%   \end{pmatrix}\,,
% &
%   \hat A & = \begin{pmatrix}
%    1 & -\frac{5}{12} \\[0.2cm]
%    0 & -\frac{1}{16}
%   \end{pmatrix}\,,
% &
%   \hat B & = \begin{pmatrix}
%   1 & -\frac{1}{6} & 0 &
%     -\frac{1}{4} \\[0.2cm]
%   0 & 0 & 0 & -\frac{1}{16} \\[0.2cm]
%   0 & \frac{1}{2} &
%     -\frac{1}{2} & \frac{1}{12}
%     \\[0.2cm]
%   0 & 0 & -\frac{1}{8} &
%     \frac{1}{48} \\
%   \end{pmatrix} \,,
% \end{align}
% and the vectors:
% \begin{align*}
%   \vec{C}_\text{BMU} & = \Big\{
%     \WcL[s]{SRR}, \WcL[s]{TRR},\;
%     \WcL[c]{SRR}, \WcL[c]{TRR}, \WcLtil[c]{SRR}, \WcLtil[c]{TRR}, \;
%     \WcL[b]{SRR}, \WcL[b]{TRR}, \WcLtil[b]{SRR}, \WcLtil[b]{TRR}\Big\} \,,
% \\
%   \vec{C}_\text{JMS} & = \Big\{
%     \wcL[S1,RR]{dd}{2122},   \wcL[S8,RR]{dd}{2122},
% \\ & \qquad
%     \wcL[S1,RR]{ud}{2221},   \wcL[S8,RR]{ud}{2221},
%     \wcL[S1,RR]{uddu}{2122}, \wcL[S8,RR]{uddu}{2122},
% \\ & \qquad
%     \wcL[S1,RR]{dd}{2133},   \wcL[S8,RR]{dd}{2133},
%     \wcL[S1,RR]{dd}{2331},   \wcL[S8,RR]{dd}{2331}\Big\} \,.
% \end{align*}

%--------+---------+---------+---------+---------+---------+---------+---------+
%
% References
%
%--------+---------+---------+---------+---------+---------+---------+---------+

\renewcommand{\refname}{R\lowercase{eferences}}

\addcontentsline{toc}{section}{References}

\bibliographystyle{JHEP}

\small

\bibliography{Bookallrefs}

\end{document}